\documentclass[twocolumn]{aastex6}
\fullcollaborationName{The Dark Energy Survey Collaboration}

\slugcomment{DES 2016-0205, FERMILAB-PUB-17-029-AE-CD}

\begin{document}


\def \sysname {DES J0408-5354}

\title{Discovery of the Lensed Quasar System \sysname}

%% Use \author, \affil, plus the \and command to format author and affiliation 
%% information.  If done correctly the peer review system will be able to
%% automatically put the author and affiliation information from the manuscript
%% and save the corresponding author the trouble of entering it by hand.
%%
%% The \affil should be used to document primary affiliations and the
%% \altaffil should be used for secondary affiliations, titles, or email.

%% Authors with the same affiliation can be grouped in a single

% Author list file generated with: mkauthlist 0+unknown 
% mkauthlist -j aastex -a DES-2016-0205_order.csv DES-2016-0205_author_list_save.csv DES-2016-0205_author_list.tex 

\def\andname{}

\author{
H.~Lin\altaffilmark{1},
E.~Buckley-Geer\altaffilmark{1},
A.~Agnello\altaffilmark{2},
F.~Ostrovski\altaffilmark{3,4},
R.~G.~McMahon\altaffilmark{3,4},
B.~Nord\altaffilmark{1},
N.~Kuropatkin\altaffilmark{1},
D.~L.~Tucker\altaffilmark{1},
T.~Treu\altaffilmark{5},
J.~H.~H.~Chan\altaffilmark{6,7},
S.~H.~Suyu\altaffilmark{7},
H.~T.~Diehl\altaffilmark{1},
T.~Collett\altaffilmark{8},
M.S.S.~Gill\altaffilmark{9},
A.~More\altaffilmark{10},
A.~Amara\altaffilmark{11},
M.~W.~Auger\altaffilmark{3},
F.~Courbin\altaffilmark{12},
C.~D.~Fassnacht\altaffilmark{13},
J.~Frieman\altaffilmark{1,14},
P.~J.~Marshall\altaffilmark{15},
G.~Meylan\altaffilmark{12},
C.~E.~Rusu\altaffilmark{16},
T. M. C.~Abbott\altaffilmark{17},
F.~B.~Abdalla\altaffilmark{18,19},
S.~Allam\altaffilmark{1},
M.~Banerji\altaffilmark{3,4},
K.~Bechtol\altaffilmark{20},
A.~Benoit-L{\'e}vy\altaffilmark{21,18,22},
E.~Bertin\altaffilmark{21,22},
D.~Brooks\altaffilmark{18},
D.~L.~Burke\altaffilmark{23,9},
A. Carnero Rosell\altaffilmark{24,25},
M.~Carrasco~Kind\altaffilmark{26,27},
J.~Carretero\altaffilmark{28},
F.~J.~Castander\altaffilmark{29},
M.~Crocce\altaffilmark{29},
C.~B.~D'Andrea\altaffilmark{30},
L.~N.~da Costa\altaffilmark{24,25},
S.~Desai\altaffilmark{31},
J.~P.~Dietrich\altaffilmark{32,33},
T.~F.~Eifler\altaffilmark{34},
D.~A.~Finley\altaffilmark{1},
B.~Flaugher\altaffilmark{1},
P.~Fosalba\altaffilmark{29},
J.~Garc\'ia-Bellido\altaffilmark{35},
E.~Gaztanaga\altaffilmark{29},
D.~W.~Gerdes\altaffilmark{36,37},
D.~A.~Goldstein\altaffilmark{38,39},
D.~Gruen\altaffilmark{23,9},
R.~A.~Gruendl\altaffilmark{26,27},
J.~Gschwend\altaffilmark{24,25},
G.~Gutierrez\altaffilmark{1},
K.~Honscheid\altaffilmark{40,41},
D.~J.~James\altaffilmark{42,17},
K.~Kuehn\altaffilmark{43},
O.~Lahav\altaffilmark{18},
T.~S.~Li\altaffilmark{1,44},
M.~Lima\altaffilmark{45,24},
M.~A.~G.~Maia\altaffilmark{24,25},
M.~March\altaffilmark{30},
J.~L.~Marshall\altaffilmark{44},
P.~Martini\altaffilmark{40,46},
P.~Melchior\altaffilmark{47},
F.~Menanteau\altaffilmark{26,27},
R.~Miquel\altaffilmark{48,28},
R.~L.~C.~Ogando\altaffilmark{24,25},
A.~A.~Plazas\altaffilmark{34},
A.~K.~Romer\altaffilmark{49},
E.~Sanchez\altaffilmark{50},
R.~Schindler\altaffilmark{9},
M.~Schubnell\altaffilmark{37},
I.~Sevilla-Noarbe\altaffilmark{50},
M.~Smith\altaffilmark{51},
R.~C.~Smith\altaffilmark{17},
F.~Sobreira\altaffilmark{24,52},
E.~Suchyta\altaffilmark{53},
M.~E.~C.~Swanson\altaffilmark{27},
G.~Tarle\altaffilmark{37},
D.~Thomas\altaffilmark{8},
A.~R.~Walker\altaffilmark{17}
\\ \vspace{0.2cm} (DES Collaboration) \\
}

\altaffiltext{1}{Fermi National Accelerator Laboratory, P. O. Box 500, Batavia, IL 60510, USA}
\altaffiltext{2}{European Southern Observatory, Karl-Schwarzschild-Strasse 2, 85748 Garching bei M\"unchen, Germany}
\altaffiltext{3}{Institute of Astronomy, University of Cambridge, Madingley Road, Cambridge CB3 0HA, UK}
\altaffiltext{4}{Kavli Institute for Cosmology, University of Cambridge, Madingley Road, Cambridge CB3 0HA, UK}
\altaffiltext{5}{Department of Physics and Astronomy, University of California, Los Angeles, CA 90095, USA}
\altaffiltext{6}{Institute of Astronomy and Astrophysics, Academia Sinica, P.O. Box 23-141, Taipei 10617, Taiwan}
\altaffiltext{7}{Max-Planck-Institut fur Astrophysik, Karl-Schwarzschild-Str.~1, D-85741 Garching, Germany}
\altaffiltext{8}{Institute of Cosmology \& Gravitation, University of Portsmouth, Portsmouth, PO1 3FX, UK}
\altaffiltext{9}{SLAC National Accelerator Laboratory, Menlo Park, CA 94025, USA}
\altaffiltext{10}{Kavli Institute for the Physics and Mathematics of the Universe (WPI), University of Tokyo Institutes for Advanced Study (UTIAS), University of Tokyo, Chiba, 277-8583, Japan}
\altaffiltext{11}{Department of Physics, ETH Zurich, Wolfgang-Pauli-Strasse 16, CH-8093 Zurich, Switzerland}
\altaffiltext{12}{Laboratoire d'Astrophysique, Ecole Polytechnique F\'ed\'erale de Lausanne (EPFL), Observatoire de Sauverny, CH-1290 Versoix, Switzerland}
\altaffiltext{13}{Department of Physics, University of California Davis, 1 Shields Avenue, Davis, CA 95616, USA}
\altaffiltext{14}{Kavli Institute for Cosmological Physics, University of Chicago, Chicago, IL 60637, USA}
\altaffiltext{15}{Kavli Institute for Particle Astrophysics and Cosmology, Stanford University, 452 Lomita Mall, Stanford, CA 94305, USA}
\altaffiltext{16}{Department of Physics, University of California, Davis, One Shields Avenue, Davis, CA 95616, USA}
\altaffiltext{17}{Cerro Tololo Inter-American Observatory, National Optical Astronomy Observatory, Casilla 603, La Serena, Chile}
\altaffiltext{18}{Department of Physics \& Astronomy, University College London, Gower Street, London, WC1E 6BT, UK}
\altaffiltext{19}{Department of Physics and Electronics, Rhodes University, PO Box 94, Grahamstown, 6140, South Africa}
\altaffiltext{20}{LSST, 933 North Cherry Avenue, Tucson, AZ 85721, USA}
\altaffiltext{21}{CNRS, UMR 7095, Institut d'Astrophysique de Paris, F-75014, Paris, France}
\altaffiltext{22}{Sorbonne Universit\'es, UPMC Univ Paris 06, UMR 7095, Institut d'Astrophysique de Paris, F-75014, Paris, France}
\altaffiltext{23}{Kavli Institute for Particle Astrophysics \& Cosmology, P. O. Box 2450, Stanford University, Stanford, CA 94305, USA}
\altaffiltext{24}{Laborat\'orio Interinstitucional de e-Astronomia - LIneA, Rua Gal. Jos\'e Cristino 77, Rio de Janeiro, RJ - 20921-400, Brazil}
\altaffiltext{25}{Observat\'orio Nacional, Rua Gal. Jos\'e Cristino 77, Rio de Janeiro, RJ - 20921-400, Brazil}
\altaffiltext{26}{Department of Astronomy, University of Illinois, 1002 W. Green Street, Urbana, IL 61801, USA}
\altaffiltext{27}{National Center for Supercomputing Applications, 1205 West Clark St., Urbana, IL 61801, USA}
\altaffiltext{28}{Institut de F\'{\i}sica d'Altes Energies (IFAE), The Barcelona Institute of Science and Technology, Campus UAB, 08193 Bellaterra (Barcelona) Spain}
\altaffiltext{29}{Institut de Ci\`encies de l'Espai, IEEC-CSIC, Campus UAB, Carrer de Can Magrans, s/n,  08193 Bellaterra, Barcelona, Spain}
\altaffiltext{30}{Department of Physics and Astronomy, University of Pennsylvania, Philadelphia, PA 19104, USA}
\altaffiltext{31}{Department of Physics, IIT Hyderabad, Kandi, Telangana 502285, India}
\altaffiltext{32}{Excellence Cluster Universe, Boltzmannstr.\ 2, 85748 Garching, Germany}
\altaffiltext{33}{Faculty of Physics, Ludwig-Maximilians-Universit\"at, Scheinerstr. 1, 81679 Munich, Germany}
\altaffiltext{34}{Jet Propulsion Laboratory, California Institute of Technology, 4800 Oak Grove Dr., Pasadena, CA 91109, USA}
\altaffiltext{35}{Instituto de Fisica Teorica UAM/CSIC, Universidad Autonoma de Madrid, 28049 Madrid, Spain}
\altaffiltext{36}{Department of Astronomy, University of Michigan, Ann Arbor, MI 48109, USA}
\altaffiltext{37}{Department of Physics, University of Michigan, Ann Arbor, MI 48109, USA}
\altaffiltext{38}{Department of Astronomy, University of California, Berkeley,  501 Campbell Hall, Berkeley, CA 94720, USA}
\altaffiltext{39}{Lawrence Berkeley National Laboratory, 1 Cyclotron Road, Berkeley, CA 94720, USA}
\altaffiltext{40}{Center for Cosmology and Astro-Particle Physics, The Ohio State University, Columbus, OH 43210, USA}
\altaffiltext{41}{Department of Physics, The Ohio State University, Columbus, OH 43210, USA}
\altaffiltext{42}{Astronomy Department, University of Washington, Box 351580, Seattle, WA 98195, USA}
\altaffiltext{43}{Australian Astronomical Observatory, North Ryde, NSW 2113, Australia}
\altaffiltext{44}{George P. and Cynthia Woods Mitchell Institute for Fundamental Physics and Astronomy, and Department of Physics and Astronomy, Texas A\&M University, College Station, TX 77843,  USA}
\altaffiltext{45}{Departamento de F\'{\i}sica Matem\'atica,  Instituto de F\'{\i}sica, Universidade de S\~ao Paulo,  CP 66318, CEP 05314-970, S\~ao Paulo, SP,  Brazil}
\altaffiltext{46}{Department of Astronomy, The Ohio State University, Columbus, OH 43210, USA}
\altaffiltext{47}{Department of Astrophysical Sciences, Princeton University, Peyton Hall, Princeton, NJ 08544, USA}
\altaffiltext{48}{Instituci\'o Catalana de Recerca i Estudis Avan\c{c}ats, E-08010 Barcelona, Spain}
\altaffiltext{49}{Department of Physics and Astronomy, Pevensey Building, University of Sussex, Brighton, BN1 9QH, UK}
\altaffiltext{50}{Centro de Investigaciones Energ\'eticas, Medioambientales y Tecnol\'ogicas (CIEMAT), Madrid, Spain}
\altaffiltext{51}{School of Physics and Astronomy, University of Southampton,  Southampton, SO17 1BJ, UK}
\altaffiltext{52}{Universidade Federal do ABC, Centro de Ci\^encias Naturais e Humanas, Av. dos Estados, 5001, Santo Andr\'e, SP, Brazil, 09210-580}
\altaffiltext{53}{Computer Science and Mathematics Division, Oak Ridge National Laboratory, Oak Ridge, TN 37831}

%% Use the \and command so offset the last author.
%%\and


%\altaffiltext{1}{AAS Journals Data Scientist}

\begin{abstract}

We report the discovery and spectroscopic confirmation of the quad-like
lensed quasar system \sysname\ found in the Dark Energy Survey (DES) 
Year 1 (Y1) data.
This system was discovered during a search for DES Y1 strong lensing 
systems using a method that identified candidates as
red galaxies with multiple blue neighbors.
\sysname\ consists of a central red galaxy surrounded by three bright 
($i < 20$) blue objects and a fourth red object.
Subsequent spectroscopic observations using the Gemini South telescope
confirmed that the three blue objects are indeed the lensed images of a 
quasar with redshift $z = 2.375$, and that the central red object is an 
early-type lensing galaxy with redshift $z = 0.597$.
\sysname\ is the first quad lensed quasar system to be found in DES
and begins to demonstrate the potential of DES to discover and dramatically 
increase the sample size of these very rare objects.

\end{abstract}

\keywords{gravitational lensing: strong --- 
quasars: general --- surveys}



\section{Introduction} \label{sec:intro}

Strong gravitational lensing systems provide valuable tools for 
studying the properties and evolution of galaxies and quasars, for
measuring the distribution of dark matter, and for constraining
cosmological parameters.
In particular, lensed quasar systems can provide information on, for example,
intervening absorption line systems \citep[e.g.,][]{smette95}, 
properties of quasar host galaxies \citep[e.g.,][]{peng06}, 
dark matter subtructure in the lens \citep[e.g.,][]{dk02},
%\citep[e.g.,][]{nierenberg14}, 
and the stellar content of the lensing galaxy \citep[e.g.,][]{schechter14}.
Moreover, when combined with time delay measurements and careful lens modeling,
lensed quasar systems can provide powerful cosmological constraints
that are complementary to those of other techniques \citep[e.g.,][]{bonvin17,suyu16,tm16}.

Previously, large samples of lensed quasars have been found by surveys
such as the Cosmic Lens All-Sky Survey \citep[CLASS;][]{myers03,browne03}
in the radio and the Sloan Digital Sky Survey Quasar Lens Search 
\citep[SQLS;][]{oguri06,inada12} and the SDSS-III BOSS quasar lens survey 
\citep{more16} in the optical.
The Dark Energy Survey \citep[DES;][]{deswhitepaper,lahav16} is an ongoing imaging
survey covering 5000~deg$^2$ of the Southern Galactic Cap in the $grizY$
filters using the Dark Energy Camera \citep{flaugher15}, 
and it holds the promise of significantly increasing the numbers
of lensed quasars.
In particular, based on the forecasts of \cite{om10}, we expect to find in
DES about 120 lensed quasar systems brighter than $i = 21$
\citep[magnitude limit applies to the fainter image for pairs and third brightest image for quadruple systems, or quads; 
see Figure~1 of][]{ostrovski17}.  
DES, specifically with the external collaboration project 
STRIDES\footnote{\url{http://strides.astro.ucla.edu/}} (PI T.\ Treu),
aims to use the resulting large lensed quasar sample for the primary
science goal of improving constraints on cosmological parameters.
This lensed quasar sample is predicted to include about 20 of the very rare
quad systems which will provide 
additional valuable information compared to pair systems
for constraining lens models 
(specifically extra constraints from 2 more positions and 2 more time delays),
in particular for cosmology purposes
\citep[e.g.,][]{suyu13} and substructure studies \citep[e.g.,][]{dk02}.

To date, we have discovered and spectroscopically confirmed three lensed quasar
pair systems \citep{agn15b,ostrovski17} in DES.
Here in this letter we 
report the discovery and confirmation of the first lensed quasar quad 
(or quad-like) system, \sysname, in the Dark Energy Survey.
We first describe our lensed quasar search procedure, system discovery, and
photometric data in \S\ref{sec:search}.
We then describe our spectroscopic observations and present our data in 
\S\ref{sec:spectra}.
We summarize and conclude in \S\ref{sec:conclusions}.
Detailed photometry analysis and lens modeling for this system are presented in a companion 
paper, \cite{agn17}, to which we will refer the reader where relevant below.

\section{Search and Discovery}\label{sec:search}

The system \sysname\ was discovered during a systematic visual search for 
candidate strong lensing systems in the Dark Energy Survey 
Year 1 (Y1) data \citep{diehl14}.
A number of different search methods have been applied to the DES Y1 data,
but the specific method involved in this case was a ``blue-near-red'' 
technique, where we first automatically identified candidate systems of
red galaxies with multiple neighboring blue objects within some radius, 
and then used visual inspections of these systems to rate them and
to select the best candidates for subsequent spectroscopic follow-up.
This method has been used previously in the Sloan Digital Sky 
Survey (SDSS) data to search for bright strong lensing systems and then 
to confirm them spectroscopically with good success \citep[e.g.,][]{diehl09,belokurov09}.

Here we started with a DES Y1 red galaxy sample selected using the redMaGiC 
technique \citep{redmagic}.
To minimize stellar contamination of this sample, we rejected objects
with {\tt SExtractor} \citep{sextractor} {\tt spread\_model}\citep{desai12,bouy13} values~$ \leq 0.01$.
We then identified as our initial set of candidates those redMaGiC galaxies
with three or more blue objects within a radius $< 10\arcsec$, where we defined
a blue object as one with colors $-1 <= g-r < 1$ and $-1 <= r-i < 1$.
We did not apply any star/galaxy separation cut to the blue objects,
but did reject objects that are saturated in any of the $g,r,i$ filters
using a {\tt SExtractor flags}~$ <= 3$ cut.
We also applied a magnitude cut $r < 22$ on the blue objects to
keep the number of candidates manageable for the visual inspection step
and to have relatively brighter candidates to ease follow-up
spectroscopic redshift measurements.

\begin{figure*}[ht!]
\plotone{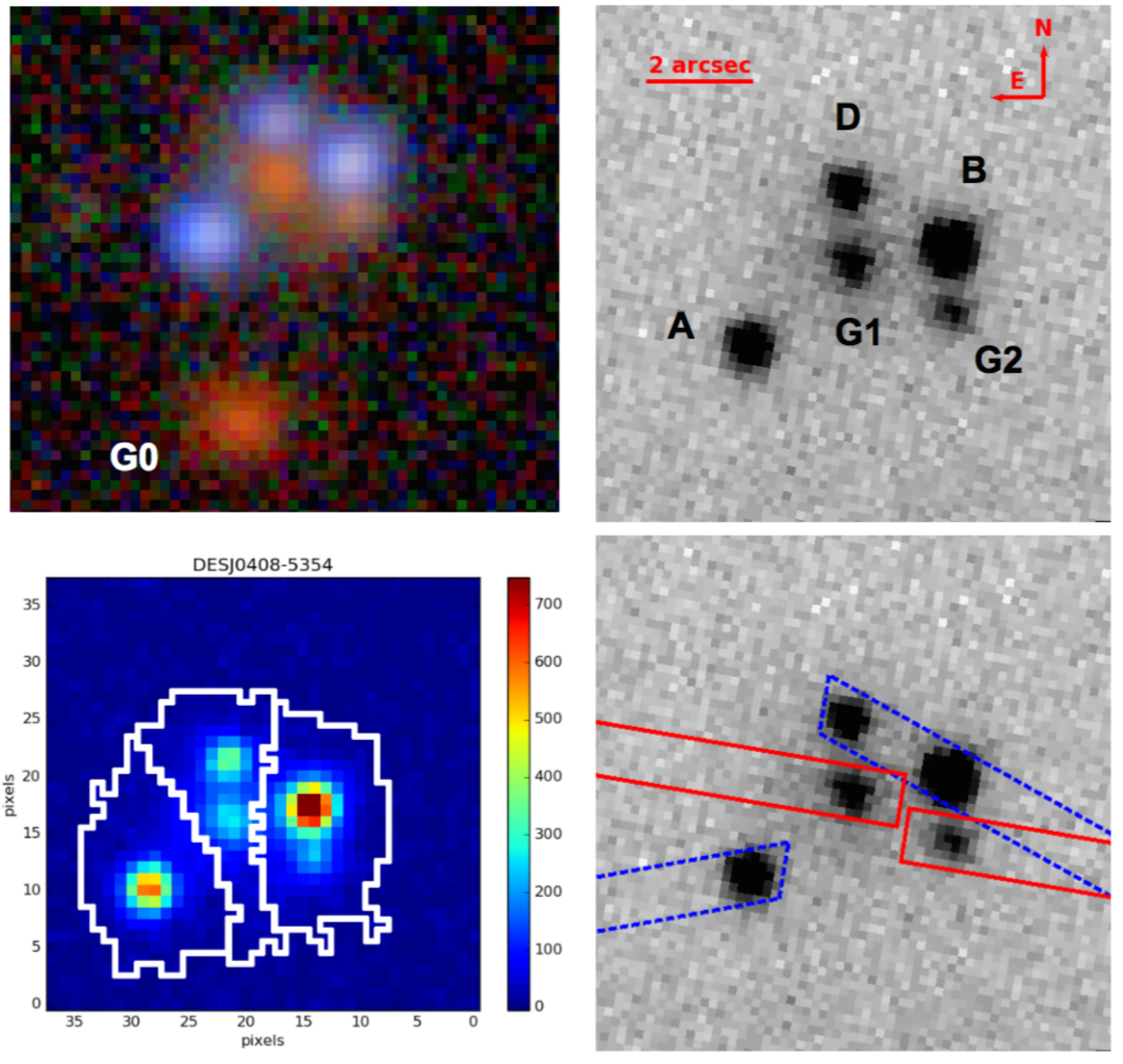}
\caption{DES $gri$ color composite discovery image (top left),
{\tt SExtractor} segmentation map plotted over the
DES $i$-band coadded image (bottom left), 
Gemini $i$-band acquisition image (top right), and the
spectroscopic slit layout (bottom right).
%North is up and east is to the left.
%A 2$\arcsec$-long horizontal red line in the right image indicates the spatial scale.
The central red lensing galaxy is G1, 
the three blue lensed quasar images are A, B, and D,
a fourth red image is G2, and the redMaGiC galaxy is G0.
The 5 components (A,B,D,G1,G2) of the system are not fully 
separated in the DES catalog, with D+G1 and B+G2 remaining blended.
\label{fig:images}}
\end{figure*}

Applying the above criteria resulted in a list of 6526 systems that one
of us (HL) then inspected visually by examining their DES $gri$ color composite
images.
The discovery image of \sysname\ is shown in Figure~\ref{fig:images},
where we see a central red galaxy (G1) surrounded by 
three blue objects (A, B, and D) and a fourth red object (G2)
\citep[objects labeled as in][]{agn17}.
This system stood out as a potential quadruply lensed quasar system,
except that the fourth putative lensed quasar image has a
conspicuously redder color than those of the other three images, 
suggesting that the fourth image may instead be that of a foregound
red galaxy, which is possibly also associated with the central red lensing
galaxy.  
Here the system would be a triple, where two of the quad's images are blended 
into one due to a fold configuration.
Another possibility is that the fourth image is a blend of a foreground red
galaxy with the fourth blue lensed quasar image, as the fourth
image in Figure~\ref{fig:images} appears not as red as the central galaxy.
Regardless, the quad-like configuration of \sysname\ made it a very 
good candidate for subsequent spectroscopy, described in the next section.
We present a detailed lensing model for this system in the companion paper
\cite{agn17}.

%object A
%select mag_auto_i, coadd_objects_id, alphawin_j2000_i, deltawin_j2000_i, ra, dec from y1a1_coadd_objects where coadd_objects_id = 3070265549;
%   MAG_AUTO_I  COADD_OBJECTS_ID  ALPHAWIN_J2000_I  DELTAWIN_J2000_I         RA        DEC
%1     19.7476        3070265549         62.091333        -53.900266  62.091333 -53.900266

%object D+G1
%select mag_auto_i, coadd_objects_id, alphawin_j2000_i, deltawin_j2000_i, ra, dec from y1a1_coadd_objects where coadd_objects_id = 3070264166;
%   MAG_AUTO_I  COADD_OBJECTS_ID  ALPHAWIN_J2000_I  DELTAWIN_J2000_I         RA        DEC
%1     19.5726        3070264166         62.090469        -53.899641  62.090469 -53.899641

%object B+G2
%select mag_auto_i, coadd_objects_id, alphawin_j2000_i, deltawin_j2000_i, ra, dec from y1a1_coadd_objects where coadd_objects_id = 3070265353;
%   MAG_AUTO_I  COADD_OBJECTS_ID  ALPHAWIN_J2000_I  DELTAWIN_J2000_I         RA        DEC
%1     19.1806        3070265353         62.089596        -53.899776  62.089596 -53.899776

\begin{figure*}
\begin{center}
\hspace*{-1.5cm}\includegraphics[width=1.2\textwidth]{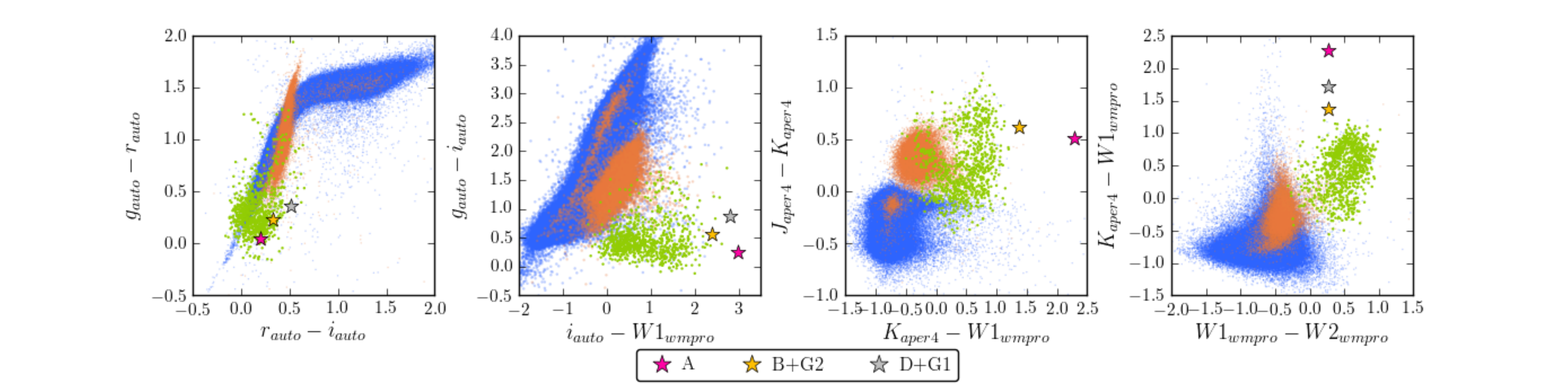}
\end{center}
\caption{Color-color plots showing the location of \sysname\ components 
A (pink stars), D+G1 (gray stars) and B+G2 (yellow stars) in different color 
spaces. 
We did not assign W1 or W2 fluxes to each individual component, so values 
show the sum of all flux in those bands.  
For comparison, the color loci of quasars (green), point sources (blue), 
and extended sources (orange) are populated by the objects present in the 
GMM training set. All magnitudes are in the AB system.}
\label{fig:quadcolor}
\end{figure*}

In the DES Y1 catalog, \sysname\ is composed of three objects due to the
fact that components D and G1 are blended and that B and G2 are also blended. 
As such, the catalog contains entries from  A, D+G1, and B+G2. 
The {\tt SExtractor} segmentation map for the system is shown in 
Figure~\ref{fig:images}, overplotted on the DES $i$-band image of the system. 
We need to point out here that all three catalog objects actually 
meet our blue object color criteria given before 
(see Figure~\ref{fig:quadcolor}, leftmost panel), 
so that the redMaGiC galaxy for this system is 
not G1 but rather the red galaxy G0 marked in Figure~\ref{fig:images}.
G0 is about $6\arcsec$ away from G1 and has a redMaGiC photometric redshift
of $0.66 \pm 0.03$, close to G1's spectroscopic redshift of 0.597 (see
\S\ref{sec:spectra}).
Therefore in this case our blue-near-red method found the system via 
another red galaxy likely associated with the lensing galaxy, rather than 
directly via the red lensing galaxy itself. 
The discovery was thus likely not coincidental, but was also not by the direct
route as intended by the method.  
In similar DES Y1 blue-near-red and related searches,
we also see
other cases in galaxy group and cluster environments where the lensing 
system is being indirectly found this way \citep[][in preparation]{diehl17}.

The coordinates and photometry of the three DES catalog objects are given in
Table~\ref{tab:phot}, while we present in \cite{agn17}
a more detailed analysis that provides the deblended positions and magnitudes
of all five components in this system.
The DES {\tt SExtractor auto} magnitudes of the three catalog objects
are listed in Table~\ref{tab:phot}, as are the near-infrared Kron magnitudes
\citep{kron80} from the VISTA Hemisphere Survey \citep[VHS;][]{mcmahon13} 
and the mid-infrared magnitudes from the Wide-field Infrared Survey Explorer 
\citep[WISE;][]{wright10}.
For the near-infrared catalog, magnitudes were obtained by cross-matching the
DES and VHS catalogs using a $1.5\arcsec$ search radius. 
Component D+G1 does not have a $J$-band magnitude value due to blending with
B+G2 in this band.  
For the mid-infrared magnitudes, the DES and WISE catalogs were cross-matched 
using a $4.0\arcsec$ search radius. 
All components of the system are blended into a single WISE source, which is 
why the same values of the $W1$ and $W2$ magnitudes are displayed in the table.

%\begin{table}[h]
\floattable
\begin{deluxetable}{cccc}
\tablecaption{\sysname\ positions (J2000 coordinates) and photometry (AB magnitudes)
\label{tab:phot}}
\tablecolumns{4}
%\tablenum{2}
\tablewidth{0pt}
\tablehead{
\colhead{} &
\colhead{A} &
\colhead{D+G1} &
\colhead{B+G2} 
}
%	\label{tab:phot}
%	\begin{threeparttable}	
%	\begin{tabular}{| l c c c |} %DES mags are auto; J, H, K are kron [AB] mags; W1, W2 are [AB]
\startdata
	   \hline
%	    & A & D+G1 & B+G2 \\ \hline
RA & 62.091333  & 62.090469 & 62.089596 \\
Dec & -53.900266 & -53.899641 & -53.899776 \\
	   \hline
		$g$ & 19.99 $\pm$ 0.01 & 20.45 $\pm$ 0.01 & 19.74 $\pm$ 0.01 \\
		$r$ & 19.94 $\pm$ 0.01 & 20.08 $\pm$ 0.01 & 19.51 $\pm$ 0.01 \\
		$i$ & 19.75 $\pm$ 0.01 & 19.57 $\pm$ 0.01  &19.18 $\pm$ 0.01 \\
		$z$ & 19.51 $\pm$ 0.02 & 19.13 $\pm$ 0.02  & 18.82 $\pm$ 0.01 \\
		$Y$ & 19.48 $\pm$ 0.07 & 19.09 $\pm$ 0.05 &  18.69 $\pm$ 0.03 \\ 
		$J$ & 19.58 $\pm$ 0.06 & -- & 18.77 $\pm$ 0.03 \\ 
		$H$ & 19.62 $\pm$ 0.08 &  18.90 $\pm$ 0.04 & 18.65 $\pm$ 0.03 \\ 
		$K$ & 19.06 $\pm$ 0.07 &  18.51 $\pm$ 0.04 & 18.15 $\pm$ 0.03 \\ 
		$W1$ & -- & 16.78 $\pm$ 0.02 & -- \\
		$W2$ & -- & 16.51 $\pm$ 0.02 & -- \\
%		$W1$ & 16.78 $\pm$ 0.02 & 16.78 $\pm$ 0.02 & 16.78 $\pm$ 0.02 \\
%		$W2$ & 16.51 $\pm$ 0.02 & 16.51 $\pm$ 0.02 & 16.51 $\pm$ 0.02 \\
%		\hline
\enddata
%\tablenotetext{a}{All magnitudes are in the AB system.}
%	\end{tabular}	
%	\begin{tablenotes}
%            \item[a] All quoted magnitudes are AB.
%    \end{tablenotes}
%	\end{threeparttable}
%	 \end{center}
%\end{table}
\end{deluxetable}

After \sysname\ was first announced to the STRIDES Collaboration as a lensed quasar 
candidate from the blue-near-red search, a number of other search methods within
STRIDES were examined and also seen to identify the system as a candidate.
These other search methods are described in more detail in \cite{agn17}.
We provide here a brief summary: 
(1) the Gaussian Mixture Model (GMM) method of \cite{ostrovski17}, which uses supervised 
machine learning in a five-dimensional optical plus infrared color space and
identified \sysname\ as a candidate pair (D+G1 not found separately in $J$ band due to blending noted above);
(2) CHITAH \citep{chan15}, which uses pixel-based automatic recognition on $grizY$ 
cutout images and identified the system as a candidate pair (not flagged as quad because fourth image G2 is too red);
and (3) the Artificial Neural Network (ANN) method of \cite{agn15a,agn15b}, 
which uses $griz$ and $W1 W2$ magnitudes and identified \sysname\ 
as a candidate extended quasar.  
Figure~\ref{fig:quadcolor} shows the positions of DES catalog objects A, B+G2, and D+G1 
in the color space used by the GMM method, illustrating how the system was flagged as a 
candidate due to the quasar-like colors of its components.

%We note that \sysname\ is also selected as a candidate following the method 
%described in \cite{ostrovski17}, when using a fainter magnitude limit of 
%$i_{auto}<20$ (19 was used in that paper). 
%The quasar-like color similarity is calculated in a five-dimensional color 
%space composed of $g-i$, $i-W1$, $J-K$, $K-W1$ and $W1-W2$, and we apply 
%supervised machine learning to select the candidates through the use of 
%Gaussian Mixture Models (GMM). 
%The GMM was implemented using astroML \citep{vanderplas12} and 
%scikit-learn \citep{pedregosa11} tools and trained on objects from the 
%Stripe 82 region of the Sloan Digital Sky Survey \citep{abazajian09}.
%The training set was separated into two classes, quasars and non-quasars, 
%and each class was modeled as a set of 10 gaussians.

%The GMM search algorithm finds 79,503 quasars of which 267 are pairs within a 
%separation of 5\arcsec.  The lack of $J$-band photometry for component 
%D+G1 means it is not present in the input sample. 
%The other two components, however, are sucessfully selected as quasar 
%candidates by the method and flagged as a pair. 
%The position each component occupies in the color space used is shown in 
%Figure~\ref{fig:quadcolor}, along with the colors of the training set objects 
%for comparison (point sources are shown in blue, extended sources in orange, 
%and quasars in green).
%It is clear that the presence of a blue object in each blended component, 
%aided by the fact the entire system is a single object in WISE, is enough 
%to make the system show indubitable quasar-like colors in every DES catalog 
%component.

\section{Spectroscopic Observations}\label{sec:spectra}

Spectroscopic observations of \sysname\ were carried out using the 
Gemini Multi-Object Spectrograph (GMOS-S) on the Gemini South telescope,
as part of a larger Gemini Large and Long Program (LLP) 
(PI E.\ Buckley-Geer; program IDs GS-2015B-LP-5, GS-2016A-LP-5)
of spectroscopic follow-up for DES strong lensing systems and 
for DES photometric redshift (photo-z) calibrations.
GMOS-S was used in multi-object spectroscopy (MOS) mode on 9 Dec 2015 UT
to take spectra of the three blue objects A, B, and D.
%These objects are labeled A, B, and D 
%\citep[following the convention in][]{agn17} in the $i$-band GMOS-S 
%acquisition image shown in Figure~\ref{fig:images} (middle).
The $i$-band GMOS-S acquisition image is shown in 
Figure~\ref{fig:images} (top right).
A single slit mask was used, with one slit for A and
a second slit for B and D together; 
see blue slits in Figure~\ref{fig:images} (bottom right).
(Another 38 slits were assigned to unrelated DES photo-z calibration targets.)
Two sets of spectra were taken: blue data using the B600 grating
(dispersion $\approx 1.0$~\AA~pixel$^{-1}$) and red data using the 
R400 grating (dispersion $\approx 1.5$~\AA~pixel$^{-1}$).
We used 4$\times$900 second exposures for cosmic ray rejection
%as well as used two central wavelengths 
%(offset by 20\AA; 2 exposures per central wavelength), in order to fill
%in the wavelength gaps in the spectra arising from spatial gaps 
%between the three CCDs comprising the GMOS-S detector. 
%We took wavelength calibration (CuAr lamp) and flatfield exposures in 
%between the science exposures.
and processed the data using the IRAF Gemini reduction package.
The seeing was $0.8\arcsec$, measured from spectra of 
mask-alignment stars in the science data.

\begin{figure*}
\plotone{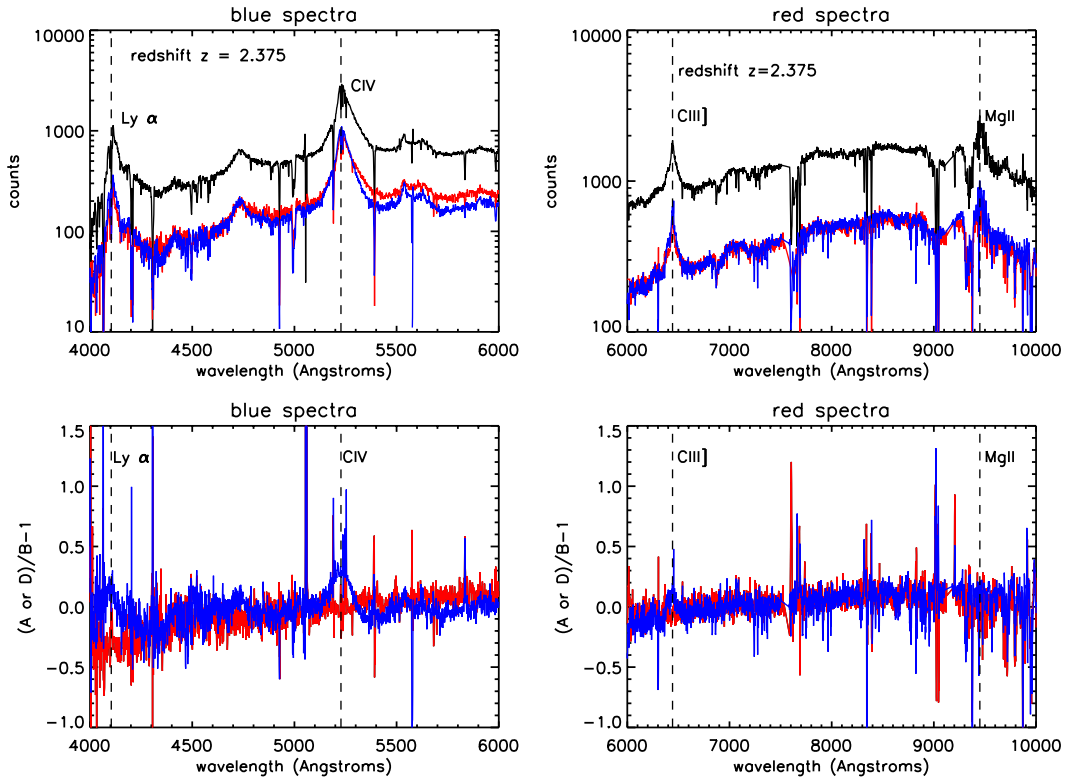}
\caption{(Top panels:) Blue (top left) and red (top right) Gemini GMOS-S spectra
for the three blue images A (red), B (black), and D (blue).
%(target labels as in Figure~\ref{fig:images}).
(Bottom panels:) Fractional differences between the blue (bottom left) and
red (bottom right) spectra of A (red) or D (blue) relative to B.  
The median counts in the spectra of A and D have been rescaled
to match those of B, over the respective blue and red wavelength ranges
shown.
\label{fig:spectra}}
\end{figure*}

From the extracted (but not flux calibrated) 1D spectra shown in 
Figure~\ref{fig:spectra} (top panels),
we clearly see that all three blue images A, B, and D show strong
quasar emission lines at the same redshift, specifically 
Ly$\alpha$ 1216\AA\ and CIV 1549\AA\ in the blue spectra, 
and CIII] 1909\AA\ and MgII 2800\AA\ in the red spectra.
From the CIII] line, which shows the cleanest, most symmetric line profile
among these four broad emission features,
we use the {\tt emsao} task in the IRAF external package {\tt rvsao}
\citep{rvsao} to report a redshift $z = 2.375$ for the quasar.
%{\color{red} Talk about lines from absorber systems.}

To compare the spectra of the three blue images in more detail, we also show
in Figure~\ref{fig:spectra} (bottom panels) the fractional differences 
between the blue and red spectra of images A and D relative to those of
image B, which has the spectra with the most counts.  
We first rescale the spectra of A and D so that they each have the same
median counts as those of B over the respective blue and red wavelength
ranges plotted. % in Figure~\ref{fig:spectra}.
The results in Figure~\ref{fig:spectra} show generally good agreement
among the spectra, especially for the red spectra.
For the blue spectra, we do see an overall slope for the spectrum of A 
relative to that of B; we attribute this slope to extinction differences 
between the lines of sight to these two images, and/or to differences in
atmospheric differential refraction between the two differently oriented
slits (Figure~\ref{fig:images}) used to observe images A and B.
In addition, we also see conspicuous differences between the spectra of D and B
at the locations of the strong Ly$\alpha$ and CIV emission lines, and to
a lesser extent at the CIII] line.
We attribute these differences to the effects of microlensing
by stars in the lensing galaxy, manifested as flux ratio differences,
in pairs of lensed quasar images, for the continuum vs.\ the emission line 
regions in the spectra \citep[e.g.,][]{motta12}.

\begin{figure*}
\plotone{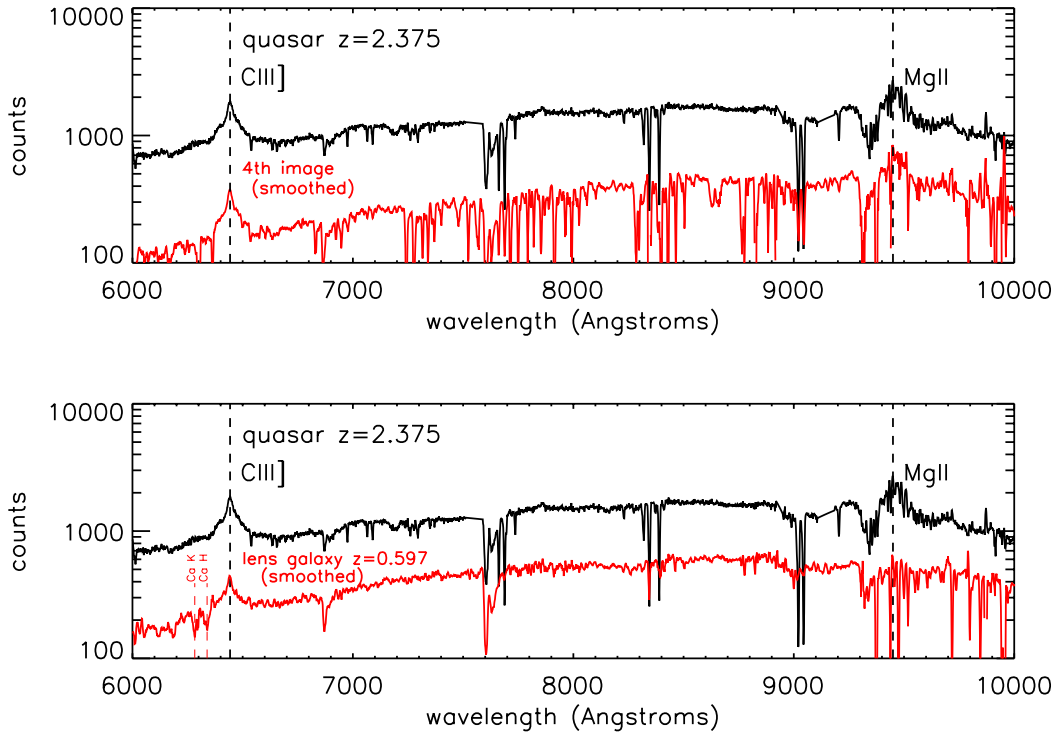}
\caption{Red Gemini GMOS-S spectra of the 
fourth image G2 (top panel, red) and of the 
central lensing galaxy G1 (bottom panel, red).  
In both panels the black spectrum is that of image B.
\label{fig:more_spectra}}
\end{figure*}

Subsequent to these observations that confirmed the three blue images as having
the same quasar redshift, we designed a second MOS slit mask targeting the
central lensing galaxy G1 and the fourth image G2,
each allocated to a single slit.  
The slits for G1 and G2 are shown in red in Figure~\ref{fig:images} (bottom right).
Red R400 observations were obtained in the Gemini South semester 2016A observing queue,
on 9 Apr 2016, under $1.1\arcsec$ seeing conditions and using the same
observing setup as above.
Blue B600 observations were also put into the queue, but no data were obtained
before the system set in 2016A.
We show the R400 spectra in Figure~\ref{fig:more_spectra}, where we see
that G2 shows broad CIII] and MgII emission at the same redshift
as found for the three blue images, while G1 shows Ca H+K absorption lines, indicating that
G1 is an early-type lensing galaxy with redshift $z = 0.597$.  

However, we also see that the G1 spectrum unexpectedly shows the same CIII] emission line 
as in the lensed quasar spectra, leading us to suspect there is contaminating light from
the neighboring lensed quasar image D, which is only about $1.3\arcsec$ away (cf.\ the $1.1\arcsec$ seeing).
Likewise, object B, the brightest quasar image, is only about $1.1\arcsec$ away from the substantially 
fainter object G2, suggesting that the CIII] line seen in the G2 spectrum is similarly contamination 
from object B. 
Using the image modeling code {\tt GALFIT} \citep{galfit}, we simulate image B as a Moffat profile 
with a FWHM of $1.1\arcsec$ and find that about 12\% of its light falls
within the 1\arcsec-wide slit and 2\arcsec-long spectral extraction aperture used for G2.
Though seemingly small, this 12\% fraction actually amounts to about 80\% of the $r$-band light from 
G2 itself, given that $r({\rm B}) = 19.95$ and $r({\rm G2}) = 21.98$ from Table~1 of \cite{agn17}.
(We use $r$ band as the CIII] line falls within that filter.)
This estimated contamination is thus substantial
and unfortunately we are unable to correct for it as image B
lies off the slit for G2 (Figure~\ref{fig:images}, bottom right) and we do not have a 
concurrent spectrum of B.
We thus cannot rule out contamination from B as the source of the CIII] line seen in the G2 spectrum
(likewise for D and G1) and will need future observations to verify whether G2 indeed
shows quasar emission features or not.

\section{Summary and Conclusions}\label{sec:conclusions}

\sysname\ was identified as a candidate lensed quasar quad system from the DES first-year
data.  Subsequent follow-up spectroscopy confirmed the three bright blue objects in
this system to be the images of a quasar at redshift $z = 2.375$, lensed by an early-type
red galaxy with redshift $z = 0.597$.
Another reddened object in the system, possibly a blend of a perturbing galaxy and the fourth 
lensed quasar image, was also observed spectroscopically, 
though without conclusive results.

Our companion paper \cite{agn17} presents a detailed model of the system, from which we
expect the longest time delay in the system to be about 80 days, making \sysname\ well
suited \citep{tm16} for time delay measurements, which we are undertaking via a 
monitoring campaign within the STRIDES Collaboration.  
The lensing model also constrains the mass of the 
possible perturbing galaxy and thus provides information about substructure in the
lensing mass distribution.
Further spectroscopic follow-up and high resolution imaging data should provide more 
needed details about the main lensing galaxy and its environment.
We thus expect this quad system to be particularly useful for the application 
of time delay cosmography \citep[e.g.][]{tm16} and substructure studies 
\citep[e.g.,][]{dk02}.
\sysname\ heralds a much larger sample of some 20 of these very rare and
valuable quad lensed quasar systems anticipated to be discovered by the Dark Energy Survey.

\acknowledgments

Funding for the DES Projects has been provided by the U.S. Department of Energy, the U.S. National Science Foundation, the Ministry of Science and Education of Spain, 
the Science and Technology Facilities Council of the United Kingdom, the Higher Education Funding Council for England, the National Center for Supercomputing 
Applications at the University of Illinois at Urbana-Champaign, the Kavli Institute of Cosmological Physics at the University of Chicago, 
the Center for Cosmology and Astro-Particle Physics at the Ohio State University,
the Mitchell Institute for Fundamental Physics and Astronomy at Texas A\&M University, Financiadora de Estudos e Projetos, 
Funda{\c c}{\~a}o Carlos Chagas Filho de Amparo {\`a} Pesquisa do Estado do Rio de Janeiro, Conselho Nacional de Desenvolvimento Cient{\'i}fico e Tecnol{\'o}gico and 
the Minist{\'e}rio da Ci{\^e}ncia, Tecnologia e Inova{\c c}{\~a}o, the Deutsche Forschungsgemeinschaft and the Collaborating Institutions in the Dark Energy Survey.

The Collaborating Institutions are Argonne National Laboratory, the University of California at Santa Cruz, the University of Cambridge, Centro de Investigaciones Energ{\'e}ticas, 
Medioambientales y Tecnol{\'o}gicas-Madrid, the University of Chicago, University College London, the DES-Brazil Consortium, the University of Edinburgh, 
the Eidgen{\"o}ssische Technische Hochschule (ETH) Z{\"u}rich, 
Fermi National Accelerator Laboratory, the University of Illinois at Urbana-Champaign, the Institut de Ci{\`e}ncies de l'Espai (IEEC/CSIC), 
the Institut de F{\'i}sica d'Altes Energies, Lawrence Berkeley National Laboratory, the Ludwig-Maximilians Universit{\"a}t M{\"u}nchen and the associated Excellence Cluster Universe, 
the University of Michigan, the National Optical Astronomy Observatory, the University of Nottingham, The Ohio State University, the University of Pennsylvania, the University of Portsmouth, 
SLAC National Accelerator Laboratory, Stanford University, the University of Sussex, Texas A\&M University, and the OzDES Membership Consortium.

The DES data management system is supported by the National Science Foundation under Grant Number AST-1138766.
The DES participants from Spanish institutions are partially supported by MINECO under grants AYA2012-39559, ESP2013-48274, FPA2013-47986, and Centro de Excelencia Severo Ochoa SEV-2012-0234.
Research leading to these results has received funding from the European Research Council under the European Union's Seventh Framework Programme FP7/2007-2013) including ERC grant agreements 240672, 291329, and 306478.

Based on observations obtained at the Gemini Observatory (acquired through the Gemini Observatory Archive and processed using the Gemini IRAF package), which is operated by the Association of Universities for Research in Astronomy, Inc., under a cooperative agreement with the NSF on behalf of the Gemini partnership: the National Science Foundation (United States), the National Research Council (Canada), CONICYT (Chile), Ministerio de Ciencia, Tecnolog\'{i}a e Innovaci\'{o}n Productiva (Argentina), and Minist\'{e}rio da Ci\^{e}ncia, Tecnologia e Inova\c{c}\~{a}o (Brazil)

The analysis presented here is based on observations obtained as part of the VISTA Hemisphere Ssurvey, ESO Programme, 179.A-2010 (PI: McMahon).

%%\allauthors


\begin{thebibliography}{}

\bibitem[Abazajian et al.(2009)]{abazajian09} Abazajian, K.~N., et al.\ 2009, \apjs, 182, 543
\bibitem[Agnello et al.(2015a)]{agn15a} Agnello, A., et al.\ 2015, \mnras, 448, 1446
\bibitem[Agnello et al.(2015b)]{agn15b} Agnello, A., et al.\ 2015, \mnras, 454, 1260
\bibitem[Agnello et al.(2017)]{agn17} Agnello, A., et al.\ 2017, \mnras, submitted
\bibitem[Belokurov et al.(2009)]{belokurov09} Belokurov, V., et al.\ 2009, \mnras, 392, 104
\bibitem[Bertin \& Arnouts(1996)]{sextractor} Bertin, E., \& Arnouts, S.\ 1996, \aaps, 117, 393
\bibitem[Bonvin et al.(2017)]{bonvin17} Bonvin, V., et al.\ 2017, \mnras, 465, 4914
\bibitem[Bouy et al.(2013)]{bouy13} Bouy, H., et al.\ 2013, \aap, 554, 101
\bibitem[Browne et al.(2003)]{browne03} Browne, I.~W.~A., et al.\ 2003, \mnras, 341, 13
\bibitem[Chan et al.(2015)]{chan15} Chan, J.~H.~H., et al.\ 2015, \apj, 807, 138
\bibitem[Dalal \& Kochanek(2002)]{dk02} Dalal, N., \& Kochanek, C.~S.\ 2002, \apj, 572, 25
\bibitem[Dark Energy Survey Collaboration(2005)]{deswhitepaper} Dark Energy Survey Collaboration, arXiv:astro-ph/0510346
\bibitem[Dark Energy Survey Collaboration(2016)]{lahav16} Dark Energy Survey Collaboration 2016, \mnras, 460, 1270
\bibitem[Desai et al.(2012)]{desai12} Desai, S., et al.\ 2012, \apj, 757, 83
\bibitem[Diehl et al.(2009)]{diehl09} Diehl, H.~T., et al.\ 2009, \apj, 707, 686
\bibitem[Diehl et al.(2014)]{diehl14} Diehl, H.~T., et al.\ 2014, Proc.\ SPIE 9149, 91490V
%\bibitem[Diehl et al.(2014)]{diehl14} Diehl, H.~T., et al.\ 2014, in Observatory Operations: Strategies, Processes, and Systems V.~p.~91490V, doi:10.1117/12.2056982
\bibitem[Diehl et al.(2017)]{diehl17} Diehl, H.~T., et al.\ 2017, in preparation
\bibitem[Flaugher et al.(2015)]{flaugher15} Flaugher, B., et al.\ 2015, \aj, 150, 150
\bibitem[Inada et al.(2012)]{inada12} Inada, N., et al.\ 2012, \aj, 143, 119
\bibitem[Kron(1980)]{kron80} Kron, R.~G.\ 1980, \apjs, 43, 305
\bibitem[Kurtz \& Mink(1998)]{rvsao} Kurtz, M.~J., \& Mink, D.~J.\ 1998, \pasp, 110, 934
\bibitem[McMahon et al.(2013)]{mcmahon13} McMahon, R.~G., Banerji, M., 
Gonzalez, E., Koposov, S.~E., Bejar, V.~J., Lodieu, N., Rebolo, R., 
VHS Collaboration 2013, The Messenger, 154, 35
\bibitem[More et al.(2016)]{more16} More, A., et al.\ 2016, \mnras, 456, 1595
\bibitem[Motta et al.(2012)]{motta12} Motta, V., Mediavilla, E., Falco, E., 
\& Mu\~{n}oz, J.~A.\ 2012, \apj, 755, 82
\bibitem[Myers et al.(2003)]{myers03} Myers, S.~.T., et al.\ 2003, \mnras, 341, 1
%\bibitem[Nierenberg et al.(2014)]{nierenberg14} Nierenberg, A.~M., et al.\ 2014, \mnras, 442, 2434
\bibitem[Oguri \& Marshall(2010)]{om10} Oguri, M., \& Marshall, P.~J.\ 2010, \mnras, 405, 2579
\bibitem[Oguri et al.(2006)]{oguri06} Oguri, M., et al.\ 2006, \aj, 132, 999
\bibitem[Ostrovski et al.(2017)]{ostrovski17} Ostrovski, F., et al.\ 2017, \mnras, 465, 4325
%\bibitem[Pedregosa et al.(2011)]{pedregosa11} Pedregosa, F., et al.\ 2011, 
%Journal of Machine Learning Research, 12, 2825
\bibitem[Peng et al.(2006)]{peng06} Peng, C.~Y., et al.\ 2006, \apj, 649, 616
\bibitem[Peng et al.(2010)]{galfit} Peng, C.~Y., et al.\ 2010, \aj, 139, 2097
\bibitem[Rozo et al.(2016)]{redmagic} Rozo, E., et al.\ 2016, \mnras, 461, 1431
\bibitem[Schechter et al.(2014)]{schechter14} Schechter, P.~L., et al.\ 2014, \apj, 793, 96
\bibitem[Smette et al.(1995)]{smette95} Smette, A., et al.\ 1995, \aaps, 113, 199
\bibitem[Suyu et al.(2013)]{suyu13} Suyu, S.~H., et al.\ 2013, \apj, 766, 70
\bibitem[Suyu et al.(2016)]{suyu16} Suyu, S.~H., et al.\ 2016, \mnras, submitted, arXiv:1607.00017
\bibitem[Treu \& Marshall(2016)]{tm16} Treu, T., \& Marshall, P.~J.\ 2016, Astronomy and Astrophysics Review, 24, 11
%\bibitem[VanderPlas et al.(2012)]{vanderplas12} VanderPlas, J., 
%Connolly, A.~J., Ivezi\'{c}, \v{Z}., \& Gray, A.\ 2012, in Proceedings of 
%Conference on Intelligent Data Understanding (CIDU), pp.\ 47-54, 
%doi:10.1109/CIDU.2012.6382200
\bibitem[Wright et al.(2010)]{wright10} Wright, E.~L., et al.\ 2010, \aj, 140, 1868

%\bibitem[Corrales(2015)]{2015ApJ...805...23C} Corrales, L.\ 2015, \apj, 805, 23
%\bibitem[Hanisch \& Biemesderfer(1989)]{1989BAAS...21..780H} Hanisch, R.~J., \& Biemesderfer, C.~D.\ 1989, \baas, 21, 780 
%\bibitem[Lamport(1994)]{lamport94} Lamport, L. 1994, LaTeX: A Document Preparation System, 2nd Edition (Boston, Addison-Wesley Professional)
%\bibitem[Schwarz et al.(2011)]{2011ApJS..197...31S} Schwarz, G.~J., Ness, J.-U., Osborne, J.~P., et al.\ 2011, \apjs, 197, 31  
%\bibitem[Vogt et al.(2014)]{2014ApJ...793..127V} Vogt, F.~P.~A., Dopita, M.~A., Kewley, L.~J., et al.\ 2014, \apj, 793, 127  

\end{thebibliography}
\end{document}